\documentclass[a4,12pt]{article}
\usepackage[dvips]{graphicx,psfrag}
\usepackage{float}
\usepackage{amsmath,amsthm,amssymb}

\def\ls{\lower0.5ex\hbox{$\buildrel >\over{\scriptstyle\sim}$}} 
\def\rs{\lower0.5ex\hbox{$\buildrel <\over{\scriptstyle\sim}$}} 
\allowdisplaybreaks
\begin{document}
\pagestyle{empty} \setlength{\footskip}{2.0cm}
\setlength{\oddsidemargin}{0.5cm}
\setlength{\evensidemargin}{0.5cm}
\renewcommand{\thepage}{-- \arabic{page} --}
\def\mib#1{\mbox{\boldmath $#1$}}
\def\bra#1{\langle #1 |}  \def\ket#1{|#1\rangle}
\def\vev#1{\langle #1\rangle} \def\dps{\displaystyle}
\newcommand{\fcal}{{\cal F}}
\newcommand{\gcal}{{\cal G}}
\newcommand{\ocal}{{\cal O}}
\newcommand{\El}{E_\ell}
\renewcommand{\thefootnote}{$\sharp$\arabic{footnote}}
\newcommand{\W}{{\scriptstyle W}}
 \newcommand{\I}{{\scriptscriptstyle I}}
 \newcommand{\J}{{\scriptscriptstyle J}}
 \newcommand{\K}{{\scriptscriptstyle K}}
%
 \def\thebibliography#1{\centerline{REFERENCES}
 \list{[\arabic{enumi}]}{\settowidth\labelwidth{[#1]}\leftmargin
 \labelwidth\advance\leftmargin\labelsep\usecounter{enumi}}
 \def\newblock{\hskip .11em plus .33em minus -.07em}\sloppy
 \clubpenalty4000\widowpenalty4000\sfcode`\.=1000\relax}\let
 \endthebibliography=\endlist
 \def\sec#1{\addtocounter{section}{1}\section*{\hspace*{-0.72cm}
 \normalsize\bf\arabic{section}.$\;$#1}\vspace*{-0.3cm}}
\def\secnon#1{\section*{\hspace*{-0.72cm}
 \normalsize\bf$\;$#1}\vspace*{-0.3cm}}
 \def\subsec#1{\addtocounter{subsection}{1}\subsection*{\hspace*{-0.4cm}
 \normalsize\bf\arabic{section}.\arabic{subsection}.$\;$#1}\vspace*{-0.3cm}}
\vspace*{-1.7cm}
\begin{flushright}
$\vcenter{
\hbox{{\footnotesize FUT and TOKUSHIMA Report}}
{ \hbox{(arXiv:1306.5387)}  }
}$
\end{flushright}

\vskip 1.4cm
\begin{center}
{\large\bf Latest constraint on nonstandard top-gluon couplings}

\vskip 0.18cm
{\large\bf at hadron colliders and its future prospect}
\end{center}

\vspace{0.9cm}

\begin{center}
\renewcommand{\thefootnote}{\alph{footnote})}
Zenr\=o HIOKI$^{\:1),\:}$\footnote{E-mail address:
\tt hioki@ias.tokushima-u.ac.jp}\ and\
Kazumasa OHKUMA$^{\:2),\:}$\footnote{E-mail address:
\tt ohkuma@fukui-ut.ac.jp}
\end{center}

\vspace*{0.4cm}
\centerline{\sl $1)$ Institute of Theoretical Physics,\
University of Tokushima}

\centerline{\sl Tokushima 770-8502, Japan}

\vskip 0.2cm
\centerline{\sl $2)$ Department of Information Science,\
Fukui University of Technology}
\centerline{\sl Fukui 910-8505, Japan}

\vspace*{1.8cm} 

\baselineskip=21pt plus 0.1pt minus 0.1pt

\centerline{ABSTRACT}

\vspace*{0.25cm}
Constraints on the nonstandard top-gluon couplings composed of 
the chromo\-magnetic- and chromo\-electric-dipole moments of the
top quark are updated by combining the latest data of top-pair
productions from the Tevatron, 7-TeV LHC, and 8-TeV LHC. We
find that adding the recent 8-TeV data to the analysis is effective
to get a stronger constraint on the chromoelectric-dipole moment
than the one from the Tevatron and 7-TeV LHC alone. We also
discuss how those constraints on the nonstandard couplings could
be further improved when the 14-TeV LHC results become available in
the near future.

\vspace*{1.5cm}

\begin{center}
\underline{Published in {\sl Phys. Rev.} {\bf D88} (2013), 017503}
\end{center}




\vfill

\setcounter{page}{0}
\newpage
\renewcommand{\thefootnote}{$\sharp$\arabic{footnote}}
\pagestyle{plain} \setcounter{footnote}{0}

The Large Hadron Collider (LHC) has discovered a new particle which seems to be
the standard-model Higgs boson~\cite{Aad:2012tfa, Chatrchyan:2012ufa}.
This discovery means the standard model is nearing completion and the LHC has
achieved one of its important aims to operate. On the other hand, however, there
have been
no positive signals suggesting the existence of new particles which are not
belonging to the framework of the standard model. That indicates that nonstandard
particles, if any, might be too heavy to be created at the present
LHC energies. Therefore, the top quark, the heaviest particle that can appear
in real experiments, is expected to play an important role in searching for new
physics beyond the standard model~\cite{Atwood:2000tu, Kamenik:2011wt}.

In this situation, an approach in terms of the effective Lagrangian composed of
only the standard-model fields is one of the most promising and general ways to
parametrize quantum effects of nonstandard particles and derive constraints
on them. Therefore, quite a number of authors have so far studied top-quark
physics at the Tevatron and LHC using this effective-Lagrangian
procedure~\cite{Atwood:1992vj}-\cite{Bernreuther:2013aga}. Among those works,
what we performed in \cite{Hioki:2009hm, HIOKI:2011xx, Hioki:2012vn} was
to combine the Tevatron and LHC data on $t\bar{t}$ productions to get a strong
restriction on possible nonstandard top-gluon couplings, i.e.,
the chromo\-magnetic- and chromo\-electric-dipole moments of the top quark.

Now that the LHC has been shut down for an upgrade to increase its colliding energy
after its successful operations at $\sqrt{s}=$7 TeV (hereafter LHC7) and $\sqrt{s}=$
8 TeV (LHC8), it will be meaningful to update those constraints by using the latest
results of the Tevatron and LHC experiments in order to clarify the current status
of new-physics search through top-gluon interactions in the effective-Lagrangian
approach. This is what we aim to perform here, which is going to be our first analysis
taking the LHC8 results into account.


The effective Lagrangian which we have adopted so far is the one proposed by
Buchm\"{u}ller and Wyler~\cite{Buchmuller:1985jz}
(see also \cite{Arzt:1994gp}-\cite{Grzadkowski:2010es}). 
In this framework, we have the following top-gluon couplings for the top-pair
productions in $pp/p\bar{p}$ collisions:
\begin{align}
{\cal L}^{\rm eff}_{ttg,ttgg}
 &= -\frac{1}{2} g_s \sum_a
  \Bigl[\,\bar{\psi}_t(x)\lambda^a \gamma^{\mu}\psi_t(x) G_\mu^a(x) \bigl.
 \nonumber\\
 &\phantom{========}-\bar{\psi}_t(x)\lambda^a\frac{\sigma^{\mu\nu}}{m_t}
  \bigl(d_V+id_A\gamma_5\bigr)
  \psi_t(x)G_{\mu\nu}^a(x)\,\Bigr], \label{eq:effint}
\end{align}
where $g_s$ is the $SU(3)$ coupling constant, and $d_V$ and $d_A$ are nonstandard
couplings corresponding to the chromomagnetic- and chromoelectric-dipole moments,
respectively. Using this Lagrangian for top-gluon interactions and
the usual standard-model Lagrangian for all the other interactions,
the total cross section of top-pair productions is derived straightforwardly
and expressed as 
\begin{equation}
\sigma(p\bar{p}/pp \to t\bar{t}X)=\sigma_{\rm SM} + \Delta \sigma (d_V, d_A),
 \label{eq:sigma}
\end{equation}
where $\sigma_{\rm SM}$ denotes the standard-model cross section and
$\Delta \sigma(d_V, d_A)$ expresses the remaining $d_{V,A}$-dependent
part.\footnote{There are also some standard-model loop effects which generate dipole
    couplings. In our preceding articles \cite{Hioki:2009hm,HIOKI:2011xx,Hioki:2012vn},
    we already have taken into account the QCD contributions to them by using
    corrected cross sections for $\sigma_{\rm SM}$ (as mentioned below), while we
    have so far neglected the electroweak (EW) contributions. It will, however, be
    required eventually to include the EW part into the analysis, which
    we discuss later before the summary.}\ 
The explicit form of Eq.(\ref{eq:sigma}) is found at the parton level in
Ref.\cite{Hioki:2009hm}. As for the parton distribution functions, we have
been using CTEQ6.6M (NNLO approximation) \cite{Nadolsky:2008zw}.

Recently, new data of top-pair productions at the Tevatron and LHC experiments were
presented by combining those from the CDF and D0 collaborations at the Tevatron, and
those from the ATLAS and CMS collaborations
at the LHC7 \cite{Tev_tt_xss,LHC_tt_xss_at_7TeV} as
\begin{alignat*}{2}
  \sigma_{\rm exp}&=7.65\pm0.41~{\rm pb} & \quad  &
     \mbox{(CDF plus D0 at the Tevatron~\cite{Tev_tt_xss})},  \\
  &=173.3\pm 10.1~{\rm pb} & \quad
     &\mbox{(ATLAS plus CMS at the LHC7~\cite{LHC_tt_xss_at_7TeV})}. 
\end{alignat*}
Furthermore, ATLAS and CMS gave new data
at $\sqrt{s}=8$ TeV \cite{ATLAS_tt_xss_at_8TeV,CMStt_xss_at_8TeV} as
\begin{alignat*}{2}
  \sigma_{\rm exp}&=241\pm32~{\rm pb} & \quad
     &\mbox{(ATLAS at the LHC8~\cite{ATLAS_tt_xss_at_8TeV})},  \\
  &=227\pm 15~{\rm pb} & \quad
     &\mbox{(CMS at the LHC8~\cite{CMStt_xss_at_8TeV})}. 
\end{alignat*}
There, the following standard-model cross sections including higher-order QCD
corrections were used for comparison based
on \cite{Cacciari:2003fi}-\cite{Kidonakis:2010dk}:
\begin{alignat}{2}
  \sigma_{\rm SM}^{\rm QCD}&=7.24^{+0.24}_{-0.27}~{\rm pb} & \quad
     &\mbox{for the Tevatron},  \nonumber\\
  &=167^{+17}_{-18}~{\rm pb} & \quad  &\mbox{for the  LHC7},\label{sigmaQCD}\\
  &=220^{+14}_{-13}~{\rm pb} & \quad  &\mbox{for the  LHC8}.\nonumber
\end{alignat}

%
%
\begin{figure}[htbp]
 \begin{minipage}{0.48\hsize}
  \begin{center}
   \includegraphics[width=70mm]{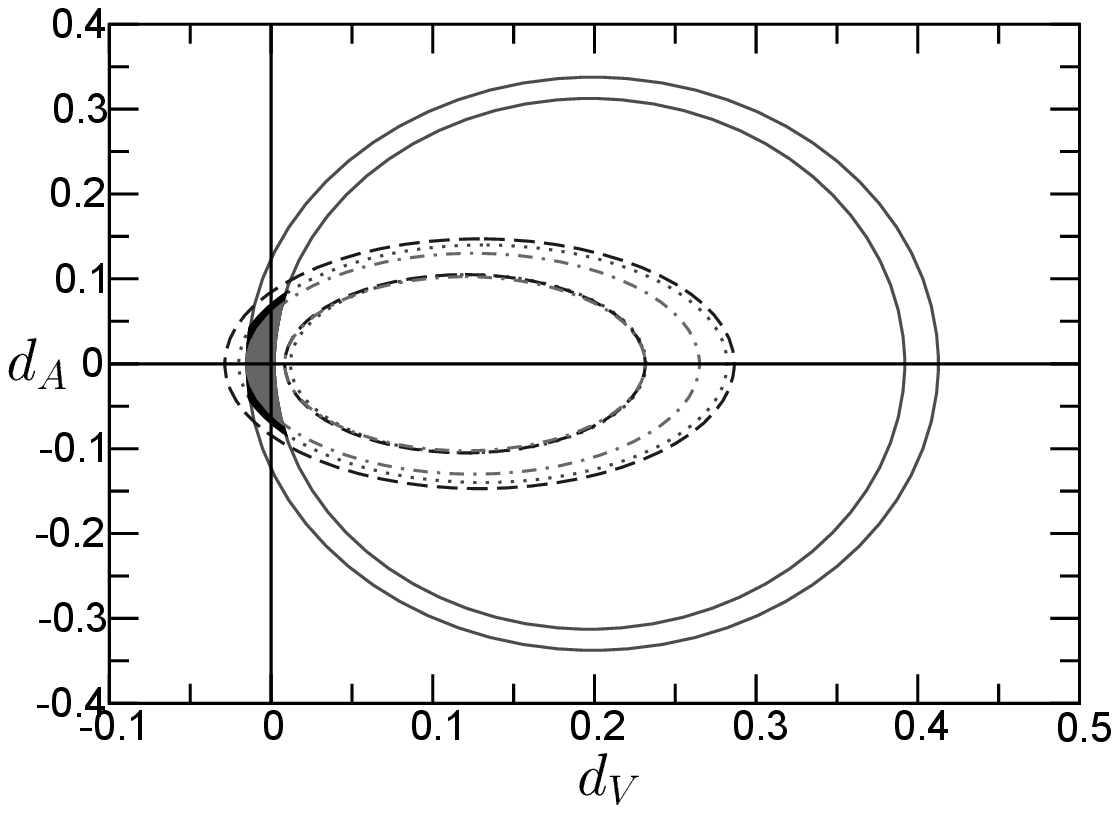}
  \caption{The $d_{V,A}$ region allowed by the Tevatron, LHC7 and LHC8 data.}
  \label{fig:current_region}
   \end{center}
 \end{minipage}
 \hspace*{0.5cm}
 \begin{minipage}{0.48\hsize}
  \begin{center}
   \includegraphics[width=70mm]{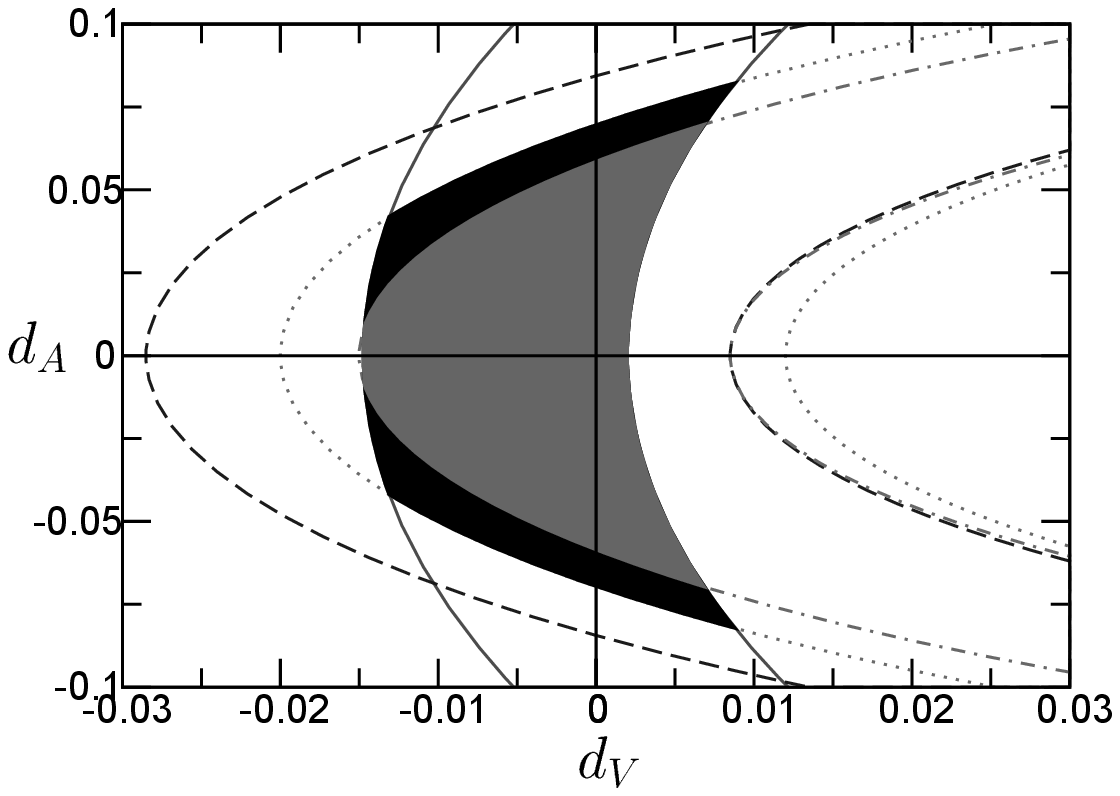}
   \caption{The enlarged view of the allowed region in Fig.\ref{fig:current_region}.}
  \label{fig:closepu_current_region}
  \end{center}
 \end{minipage}
\end{figure}

Treating the center values of these $\sigma_{\rm SM}^{\rm QCD}$ as $\sigma_{\rm SM}$ 
in Eq.(\ref{eq:sigma}), and comparing the whole $\sigma$  with the above experimental data,
the resultant allowed regions of $d_V$ and $d_A$ are shown in
Figs.\ref{fig:current_region} and \ref{fig:closepu_current_region}.
In those figures, the ellipsoidal regions given by the solid curves, the
dotted curves, the dashed curves and the dash-dotted curves are those from
the combined Tevatron (CDF \& D0), combined LHC7 (ATLAS [7 TeV] \& CMS[7 TeV]),
ATLAS [8 TeV] and CMS [8 TeV] data.\footnote{Note that readers might encounter
    a similar figure in which, however, all the curves got turned over about
    $d_V=0$ (i.e., the $d_A$ axis) as if they had performed the analysis
    with opposite-sign $d_V$.
    In that case, compare not only the sign of their nonstandard coupling but
    also the one of their standard coupling with ours, since the interference
    between these two contributions would make this seeming difference.
    Concerning $d_A$, on the other hand, any difference does not appear because
    there is no such interference and the leading term is proportional
    to $d_A^2$. This is why all the curves in the figures are symmetric about
    the $d_V$ axis.}\
In addition, the shaded regions mean the common $d_{V,A}$ regions allowed by all
the data. More specifically, the part allowed by the combined Tevatron and LHC7
is the whole shaded, i.e., the black plus gray regions, while the gray part is
the allowed one estimated by the Tevatron and LHC7 data $plus$ the ATLAS and CMS data
at $\sqrt{s}$ = 8 TeV.
This shows that the black region was excluded by taking into account the LHC8 data.
Although $d_V$ is still mainly limited by the Tevatron
data in the positive region,
the latest Tevatron and LHC data are becoming dominant in its negative region.

Since the LHC with $\sqrt{s}=$ 14 TeV (LHC14) is planned to operate in the
near future, the constraint of $d_V$ and $d_A$ could be much
more improved. In order to estimate what improvement
is expected with the increasing colliding energy, we also perform a virtual
analysis according to the above method for the LHC14. Let us use the following theoretical
prediction on the top-pair productions for $m_t$ = 173 GeV~\cite{Kidonakis:2010dk},
assuming 10\% and 5\% errors~\footnote{We mean those errors as combinations of
    the theoretical one in $\sigma_{\rm SM}$ and experimental one.}\
as the virtual-experimental value:
\begin{alignat*}{2}
  \sigma(\sqrt{s}=14\,{\rm TeV}) &=920~\pm 92 {\rm ~pb} & \quad  &\mbox{(10\% error case)},\\
         &=920~\pm 46 {\rm ~pb} & \quad  &\mbox{(5\% error case)}.
  \end{alignat*}

\begin{figure}[htbp]
  \begin{center}
   \includegraphics[width=70mm]{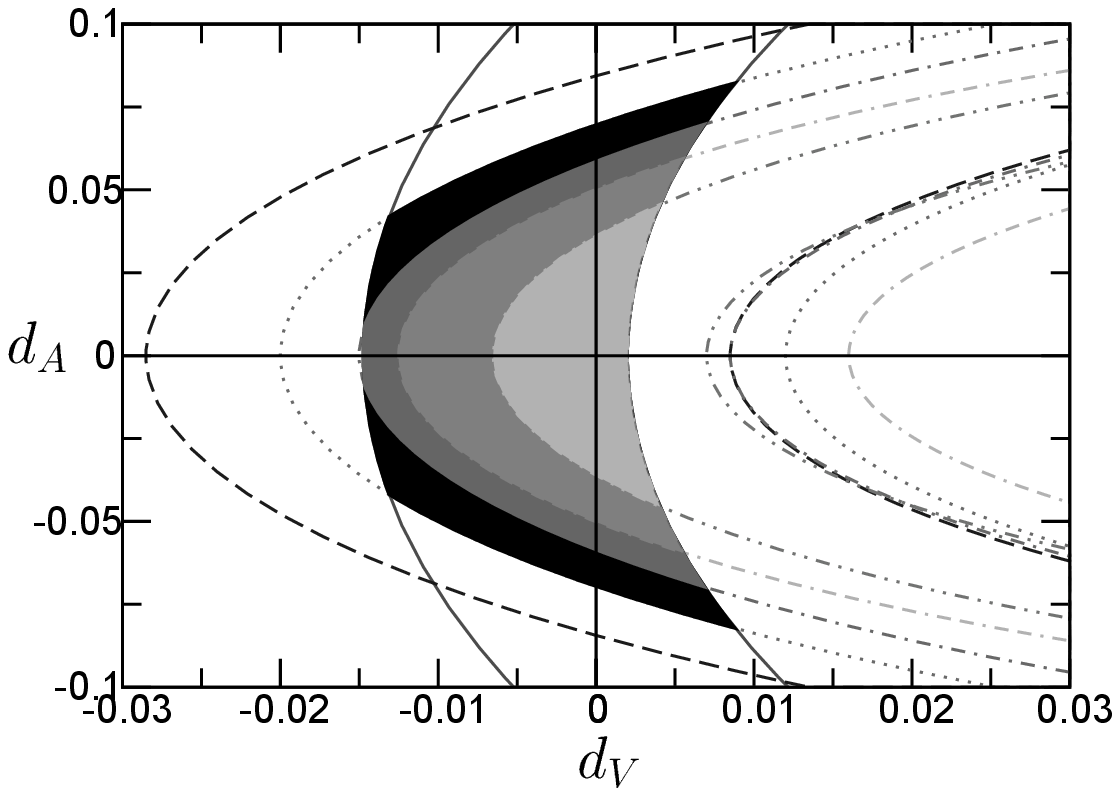}
  \caption{The expected  allowed regions of $d_{V,A}$  
  when the LHC14 data become available.}
 \label{fig:future_region}
   \end{center}
\end{figure}

The results of this virtual analysis are shown in Fig.\ref{fig:future_region}.
There, the dash-dot-dashed curves and the dot-dash-dotted curves, which 
indicate the allowed region estimated from the 10\%  and 5\% error cases, respectively,
are added to Fig.\ref{fig:closepu_current_region}. 
Moreover, the allowed regions combining the current constraint derived here and
constraints from the 10\% and 5\% error cases are described
as the middle-lighter and lighter gray regions.
As seen in Fig.\ref{fig:future_region}, the LHC14 has a potential to strengthen both of
the current individual constraints on $d_V$ and $d_A$ about twice, i.e., the allowed area
could become almost quarter its size, if the errors are controlled at about the 5\% level.

%
%

Finally, let us briefly discuss standard-model loop effects. Within
the standard model, a top-quark chromoelectric-dipole-moment term can
only arise at three-loop level through $C\!P$-violating electroweak
interactions and its contribution may be safely neglected. On the other hand,
a chromomagnetic-dipole-moment term is generated at one-loop level as
both QCD corrections (denoted as $\Delta d_V^{\rm QCD}$) and
EW corrections ($\Delta d_V^{\rm EW}$). Here, we used
the standard-model cross sections including higher-order QCD
corrections $\sigma_{\rm SM}^{\rm QCD}$ presented in Eq.(\ref{sigmaQCD})
for $\sigma_{\rm SM}$ in Eq.(\ref{eq:sigma}), that is, the $\Delta d_V^{\rm QCD}$
contribution is included in $\sigma_{\rm SM}$ implicitly with other QCD corrections,
while we have not taken account of $\Delta d_V^{\rm EW}$. Therefore, strictly
speaking, the constraint on $d_V$ shown in our figures should be understood as
the one not on $d_V$ alone but on
\begin{equation}
d_V + \Delta d_V^{\rm EW}.
\end{equation}
At present, this does not cause any serious problem, considering the size of
$\Delta d_V^{\rm EW}$ and the precision of our analysis:
According to Ref.\cite{Martinez:2007qf}, $|\Delta d_V^{\rm EW}|=9.4 \times 10^{-4}$
(for $m_{\rm Higgs}=$120 GeV), which moves the origin of the $d_V$ axis in our
figures by only about 0.001. We, however, will have to take it into account
carefully in the near future when the LHC14 starts,
which our virtual analysis is telling us.

In summary, using the latest data of top-pair productions at the Tevatron
and LHC, the current bound on the chromo\-magnetic-dipole moment ($d_V$)
and chromo\-electric-dipole moment ($d_A$) of the top quark was updated.
Although the main contribution to constraining $d_V$, especially in its
positive region, still comes from the Tevatron data, the LHC data are now
giving almost the same constraint as the Tevatron in the negative region.
For constraining $d_A$, on the other hand, the LHC8 data, which were taken
into account for the first time here, were effective to exclude some area allowed
by the Tevatron and LHC7 data alone. In addition, it was pointed out via
a virtual analysis that the current allowed area on the $d_V$-$d_A$ plane
could get almost quarter the size if the errors were controlled at the 5\% level
for the measured $t\bar{t}$ cross section at the LHC with $\sqrt{s}$ = 14 TeV.

%
\secnon{Acknowledgments}
%
This work was partly supported by the Grant-in-Aid for Scientific Research 
No. 22540284 from the Japan Society for the Promotion of Science.
Part of the algebraic and numerical calculations were carried 
out on the computer system at Yukawa Institute for Theoretical
Physics (YITP), Kyoto University.

\baselineskip=19pt plus 0.1pt minus 0.1pt

\newpage 

\end{document}